# Large Negative Magnetoresistance in Antiferromagnetic Gd$_2$Se$_3$

Santosh Karki Chhetri[1], Gokul Acharya[1], David Graf[2], Rabindra Basnet[1,3], Sumaya Rahman[4], M.M. Sharma[1], Dinesh Upreti[1], Md Rafique Un Nabi[1,6], Serhii Kryvyi[4], Josh Sakon[5], Mansour Mortazavi[3], Bo Da[7], Hugh Churchill[1,4,6], Jin Hu[1,4,6*]

[1]Department of Physics, University of Arkansas, Fayetteville, AR 72701, USA

[2]National High Magnetic Field Lab, Tallahassee, FL 32310, USA

[3]Department of Chemistry & Physics, University of Arkansas at Pine Bluff, Pine Bluff, Arkansas 71603, USA

[4] Material Science and Engineering Program, Institute for Nanoscience and Engineering, University of Arkansas, Fayetteville, AR 72701, USA

[5]Department of Chemistry & Biochemistry, University of Arkansas, Fayetteville, AR 72701, USA

[6]MonArk NSF Quantum Foundry, University of Arkansas, Fayetteville, Arkansas, USA

[7]Center for Basic Research on Materials, National Institute for Materials Science, Tsukuba, Ibaraki 305-0044, Japan

Abstract

Rare earth chalcogenides provide a great platform to study exotic quantum phenomena such as superconductivity and charge density waves. Among various interesting properties, the coupling between magnetism and electronic transport has attracted significant attention. Here, we report the investigation of such coupling in $α$-$Gd_2Se_3$ single crystals through magnetic, calorimetric, and transport property measurements. $α$-$Gd_2Se_3$ is found to display an antiferromagnetic ground state below 11 K with metamagnetic spin-flop transitions. The magnetic fluctuations remain strong above the transition temperature. Transport measurements reveal an overall metallic transport behavior with a large negative magnetoresistance of ~ 65% near the magnetic transition temperature, together with positive MR near the field-induced spin-flop transitions, which can be understood in terms of the suppression of spin scattering by the magnetic field.

jinhu@uark.edu

## I. INTRODUCTION

Rare earth chalcogenides display a variety of stoichiometric compositions such as $RX$, $RX_2$, $RX_3$, $R_3X_4$, and $R_2X_3$, where $R$ represents rare earth and $X$ represents chalcogen S, Se, and Te [1]. Rare earth chalcogenides have attracted significant interest because of their unique electronic, magnetic, optical, thermoelectric, and topological properties [2–9], which are arising from or related to the $4f$ electrons of rare earth [10,11]. In addition, tunable band gap, strong photoluminescence, and efficient luminescent properties make them promising candidates for next-generation lighting and display technologies [12–14].

Rare earth monochalcogenide $RX$ crystallizes in a NaCl-type structure [15,16]. Under high pressure, $RX$ undergoes a structural phase transition to a CsCl-type [17–21], which is accompanied by a transition from semiconducting-like to metal-like transport properties. It has been predicted that when reducing the dimensionality to the 2D limit, i.e., single atomic layer, some rare earth monochalcogenides such as Tm$X$ and Yb$X$ possess a honeycomb hexagonal lattice and display piezoelectricity [22]. In addition, for other group-III monochalcogenides Sc$X$ and Y$X$, their single layer has also been predicted to distort to a wrinkled structure, which can lead to Dirac points and nodal lines near the Fermi level [23].

For rare earth di- and tri-chalcogenides $RX_2$ and $RX_3$, they possess layered structures formed from the stacking of chalcogen $X$ and rare earth-chalcogen $R$-$X$ layers. In $RX_2$, the structure is characterized by alternative stacking of single $X$ and single $R$-$X$ layers, while the stacking of double $X$ and single $R$-$X$ layers forms $RX_3$ [5]. Interestingly, the lattice is tunable by vacancies in the chalcogen layers. Despite of an overall layered structure, various tetragonal, orthorhombic, triclinic, and monoclinic structural variations have been reported [24–26]. Among various $RX_2$ and $RX_3$, the telluride compounds have been studied extensively because of their $4f$ magnetism [25,27,28] and diverse properties such as large negative magnetoresistance [29], charge density wave [30–34] and pressure- or doping-induced superconductivity [34–39].

Compared to mono-, di-, and tri-chalcogendies, sesqui-chalcogenides $R_2X_3$ are less explored. The reported studies are mainly focused on sulfides $R_2S_3$, for which diverse crystal structures such as orthorhombic, tetragonal, cubic, monoclinic, and rhombohedral (usually denoted by α, β, γ, δ, and ε respectively) have been identified [40–49]. Many orthorhombic α-$R_2S_3$ have been reported to show single or multiple antiferromagnetic (AFM) transitions [50–54] except for α-$Sm_2S_3$ that show weak ferromagnetism at low temperatures [55]. For sesqui-selenides, the earlier studies were mainly focused on crystal structures [56,57], thermoelectric and optical properties [3,9]. For tellurides $R_2Te_3$, in addition to the thermoelectric properties [9], a recent work has revealed an AFM semimetal state in orthorhombic α-$Gd_2Te_3$ [58].

In this work, we extend the study to $Gd_2Se_3$ owing to the possible strong magnetism from the half-filled *f*-orbital of $Gd^{3+}$, which may interplay with other degrees of freedom and give rise to exotic properties such as the gigantic isotropic magnetoresistance and insulator-to-metal transition [59]. The cubic $Gd_2Se_3$ crystallizing in a defect $Th_3P_4$-type structure (due to Gd vacancies) has been reported to be a semiconductor [9,60,61]. To explore the interaction between magnetism and electronic transport, we switch to a different structure type, i.e. the orthorhombic α-phase, because the previous studies revealed a semiconductor state in α-$Gd_2S_3$ [50] but the semi-metal phase in α-$Gd_2Te_3$ [58]. Indeed, we have discovered an AFM ground state and metallic transport properties in α-$Gd_2Se_3$. More interestingly, this material displays a large negative magnetoresistance, which evidences strong coupling between magnetism and transport and can be attributed to the suppression of strong spin scattering under a magnetic field.

## II. EXPERIMENT

The $Gd_2Se_3$ single crystals used in this work were prepared by a two-step chemical vapor transport (CVT) method. First, the precursor for CVT was prepared by heating the mixture of Gd, Sb, and Se elemental powders with a ratio of 1:1:1 in a vacuum-sealed evacuated quartz tube at 850 °C for 2 days. Then, the precursor was used as a source for CVT, which were performed with a temperature gradient from 1075 °C and 975 °C for two weeks. Millimeter-size single crystals with metal luster can be obtained, as shown in the inset of Fig. 1(a). The composition and structure analyses by energy-dispersive X-ray spectroscopy (EDS) and X-ray diffraction (XRD), respectively, have revealed an orthorhombic phase of $Gd_2Se_3$. It is worth noting that the addition of Sb in the precursor is necessary to produce the desired sesqui-chalcogenides phase, otherwise, dichalcogenide $GdSe_2$ is produced. Nevertheless, the obtained crystals were found to be in a pure phase without any trace of Sb according to EDS or XRD. Temperature dependent magnetization was measured using a 7 T Magnetic property measurement system (MPMS3, Quantum Design). Field dependent magnetization up to 9 T, electronic transport using a four-probe contact configuration, and heat capacity were measured using a Physical property Measurement system (PPMS DynaCool, Quantum Design). The high field magnetoresistance measurements up to 31 T were performed at the National High Magnetic Field Laboratory (NHMFL).

## III. RESULTS AND DISCUSSION

Rietveld refinement of the XRD pattern for $Gd_2Se_3$ [Fig. 1(a)] resolves an orthorhombic lattice structure with a space group *Pnma* (i.e. *α*-phase), as shown in Fig. 1(b). The extracted structural parameters are presented in Table I. To uncover the magnetic properties of *α*-$Gd_2Se_3$, we have performed the temperature [$\chi(T)$] and field [$M(H)$] dependent magnetization measurements. According to single crystal

XRD, the natural cleavage plane of the single crystal is (201), therefore the magnetic field was applied perpendicular [$H\perp(201)$] and parallel [$H//(201)$] to the (201) plane for magnetization measurements. The temperature dependence of susceptibility ($\chi$) measured with 0.1 T magnetized field applied perpendicular [$H\perp(201)$] and parallel [$H//(201)$] to the (201) plane reveals clear magnetic transitions at 11 K, as shown in Figs. 2(a) and 2(b). Here we denote susceptibility measured under $H\perp(201)$ and $H//(201)$ as $\chi_\perp$ and $\chi_{//}$, respectively. The very weak irreversibility between zero-field-cooling (ZFC) and field-cooling (FC) measurements at very low temperatures ($T < 4$ K) in both $\chi_\perp$ and $\chi_{//}$ [Fig. 2(b), inset] suggests an AFM ground state, which is further supported by field dependence of magnetization as will be shown later. The weak irreversibility might be attributed to weak ferromagnetism arising from magnetic fluctuations due to the strong competition between AFM and ferromagnetic (FM) interactions, which has been theoretically revealed for the case of α-Gd$_2$Te$_3$ [58]. Such low-temperature magnetic fluctuation can be important in electronic transport because it can act as a source for charge carrier scattering and affect transport significantly. The Néel temperature ($T_N$) of the AFM transition shifts to a lower temperature upon increasing the magnetic field. Eventually, $T_N$ becomes unobservable down to $T = 2$ K above 5 T and 3 T fields for perpendicular i.e., $H\perp(201)$ [Fig. 2(a)] and parallel i.e., $H//(201)$ [Fig. 2(b)] directions, respectively. Such field suppression of $T_N$ is also observed in the sulfide and telluride sibling compounds α-Gd$_2$S$_3$ [62] and α-Gd$_2$Te$_3$ [58] as well as other rare earth materials [63,64]. The low field (0.1 T) $\chi_{//}$ exhibits a much sharper drop below $T_N$ as compared to $\chi_\perp$ [Figs. 2(a and b)], suggesting that the magnetic easy axis may be within or almost within the (201) plane.

In the paramagnetic (PM) state, the inverse of susceptibility $1/\chi(T)$ displays a linear temperature dependence well above $T_N$ ($T > 150$ K) [Fig. 2(a), inset], which can be described by the modified Curie-Weiss (CW) model $\chi = \chi_0 + C/(T - \theta_{cw})$, where $\chi_0$, $C$, and $\theta_{cw}$ represent the temperature-independent susceptibility, Curie constant, and Weiss temperature, respectively. The CW fitting yielded an effective moment $\mu_{\text{eff}} = \sqrt{\frac{3k_B C}{N_A}}$ of 8.34$\mu_B$, where $N_A$ is the Avogadro's number and $k_B$ is the Boltzmann constant. Such a value is very close to the theoretical moment of 7.93$\mu_B$ for Gd$^{3+}$ ion with a 4$f^{\,7}$ configuration and consistent with other reported compounds containing Gd$^{3+}$ such as the sibling compound α-Gd$_2$Te$_3$ [58] and other Gd-based compounds such as GdPS [59]. Furthermore, the fitting yields a positive $\theta_{\text{cw}} \approx 1.168$ K, which appears to be inconsistent with the AFM ground state of α-Gd$_2$Se$_3$ but might be explained by the weak ferromagnetic fluctuation due to competing AFM and FM interactions [65,66] as mentioned above. Further experimental and theoretical efforts are needed to determine the magnetic structure and clarify the enhanced $\mu_{\text{eff}}$ and positive $\theta_{\text{cw}}$ in α-Gd$_2$Se$_3$.

The AFM ground state is further supported by the isothermal magnetization measurements, which display a linear field dependence near zero fields at temperatures below $T_N \approx 11$ K, as shown in Figs. 2(c) and 2(d). Additional features can be observed at high fields. For $H\perp(201)$, the magnetization tends to saturate above 5 T at $T = 2$ K [Fig. 2(c)]. A similar tendency is also seen when $H//(201)$ [Fig. 2(d)]. As will be shown later, our high field magnetotransport reveal a complete saturation under $H\perp(201)$ around 16 T near 2 K. Though perfect moment saturation may not be achieved in magnetization measurements up to 9 T, the moment reaches 7.48 and 7.36 $\mu_B$ per Gd for $H//(201)$ and $H\perp(201)$ respectively, indicating strong polarization of the Gd moments. In addition to moment polarization under high field, one striking feature in magnetization is the three metamagnetic transitions at lower fields, which are present for $H//(201)$ but absent for $H\perp(201)$, as indicated by black arrows in Fig. 2(d). Interestingly, magnetic hysteresis is observed for these metamagnetic transitions around 0.85 T, 2.35 T, and 3.55 T at $T = 2$ K, as shown in Fig. 2(e). This implies the development of magnetic domains that might be associated with FM correlations from the canted moment, which is likely caused by the competition between AFM and FM interactions as stated above. These transitions resemble spin-flop (SF) transitions in AFM material such as MnPS$_3$ and NiPS$_3$ [67,68], which has also been observed in α-Gd$_2$Te$_3$ [58]. Because SF transitions in AFM materials are caused by the moments rotation which is driven by the magnetic field parallel to the magnetic easy axis, our observations indicate that the magnetic easy axis for α-Gd$_2$Se$_3$ is within or close to the (201) plane, which is consistent with that observed from temperature dependent susceptibility measurements as discussed above. A similar scenario has also been observed in α-Gd$_2$S$_3$ [62] and α-Gd$_2$Te$_3$ [58]. It is not clear how can multiple SF transitions occur in one material. Possible mechanisms include the presence of multiple magnetic lattices, or complicated magnetic structures with non-collinear moments. More direct experimental probes such as neutron scatterings are needed to clarify the nature of these metamagnetic transitions.

The comparison of various α-Gd$_2$(S,Se,Te)$_3$ sesqui-chalcogenides provides some insight into the nature of magnetism in those materials. In addition to α-Gd$_2$Se$_3$, α-Gd$_2$S$_3$ also displays multiple metamagnetic transitions at lower fields and spin polarization around 11 ~ 12 T [62] while α-Gd$_2$Te$_3$ exhibits only one spin flop transition without moment polarization up to 9 T field [58]. Such differences might be attributed to the nature of the AFM ground states in this family of materials. Antiferromagnetism in α-Gd$_2$Te$_3$ has been predicted to be stabilized mainly by the $4f$ Gd$^{3+}$ – $5p$ Te$^{2-}$ – $4f$ Gd$^{3+}$ super-exchange interactions [58]. A similar scenario can be expected in α-Gd$_2$S$_3$ and α-Gd$_2$Se$_3$, where magnetism could be governed by the $4f$ Gd$^{3+}$ – $5p$ (S or Se)$^{2-}$ – $4f$ Gd$^{3+}$ super-exchange interactions. The dominant role of Gd$^{3+}$ – $X^{2-}$ – Gd$^{3+}$ ($X$ = S, Se, or Te) super-exchange interaction is supported by the variation of $T_N$ magnitude that systematically increases from $T_N \approx 10$ K in α-Gd$_2$S$_3$ [54,62] to $T_N \approx 11$ K in α-Gd$_2$Se$_3$ (this work) and to $T_N \approx 15$ K in α-Gd$_2$Te$_3$ [58], which can be explained by the enhanced super-exchange interaction due to stronger orbital overlap with expanded $p$-orbitals from S to Se and to Te. Therefore, with enhanced super-exchange, α-Gd$_2$Te$_3$ possesses a more robust AFM ground state and thus needs higher field to induce moment reorientation and FM polarization.

Heat capacity measurements also provide useful information about magnetism. As shown in Fig. 2(f), a broad heat capacity peak centered at 11 K is consistent with the AFM transition temperature in the susceptibility measurements. With the application of the magnetic field, the heat capacity peak is suppressed to lower temperatures, which agrees well with the field suppression for AFM transition seen in susceptibility measurements [Figs. 2(a) and 2(b)]. Interestingly, heat capacity in the PM state is enhanced strongly by the magnetic field, as indicated by the arrow in Fig. 2(f). Such enhancement is distinct from some other rare earth-based AFM materials such as LnSnGe (Ln = Gd, Tb, and Er) [69] and SmSbTe [70] indicating very strong magnetic correlations in the PM state, which is also supported by the observation of spin polarization above $T_N$ [Figs. 2(c) and 2(d)]. Such strong magnetic correlations above $T_N$ are also probed in transport measurements, as will be discussed below.

With the characterization of magnetism, the interplay of magnetism and transport can be revealed by magnetotransport measurements. As shown in Fig. 3(a), the temperature dependence of resistivity displays overall metallic behavior showing decreased resistivity upon cooling. Over the entire temperature range (2 – 300 K), the resistivity is in the order of 1 mΩ cm, implying that $\alpha$-Gd$_2$Se$_3$ might not be a good metal. Such resistivity value is comparable to $\alpha$-Gd$_2$Te$_3$ which has been proposed to be a semimetal [58], whereas the sulfide compound $\alpha$-Gd$_2$S$_3$ is a semiconductor [54]. In addition to the orthorhombic $\alpha$-phase studied in this work, the cubic phase has been more extensively investigated, which can display both metallic and non-metallic transport behavior depending on the Gd vacancies [9,60,71]. For our $\alpha$-Gd$_2$Se$_3$, at zero field, a sharp resistivity peak at $T_N \sim$ 11 K can be observed. The resistivity peak is suppressed with the application of a perpendicular magnetic field [$H\perp(201)$] and vanishes when $\mu_0H \geq 7$ T, consistent with the suppression of $T_N$ seen in magnetic susceptibility and heat capacity measurements mentioned above. Furthermore, above 130 K, resistivity displays a linear temperature dependence that is not affected by the magnetic field, as shown in the inset of Fig. 3(a).

The suppression of the resistivity peak by magnetic field leads to remarkable negative magnetoresistance (MR), which can be better visualized in the field dependence for resistivity as shown in Fig. 3(b). Here the MR is normalized to the zero-field resistivity value, i.e., MR = $\frac{\rho(H)-\rho(0)}{\rho(0)}$. With this definition, large MR ~65% can be observed at 10 K and 9 T. At $T = 2$ K, MR is reduced to 54%, with a tendency toward saturation approaching 9 T. With extending the magnetic field to 31 T at the national high magnetic field lab (NHMFL), complete MR saturation reaching 57% can be achieved around 16.7 T at 2.2 K. Lower temperature to 1.5 K, the saturation MR is reduced to 49% around 12.5 T, as shown in the inset of Fig. 3(b). The temperature dependence of MR magnitude at 9 T field ($MR_{9T}$) is summarized in Fig. 3(c), from which the maximum MR near $T_N$ (= 11 K) is clearly seen. Above $T_N$, MR gradually reduces with rising temperature, reaching ~12% at 50 K and becoming hardly observable above 200 K. Those observations are reproducible in multiple samples, as evidenced by the consistent results obtained from the two different samples for the low field [Fig. 3(b), main panel] and high field [Fig. 3(b), inset] measurements.

The substantial negative MR in $\alpha$-$Gd_2Se_3$ near $T_N$ has also been observed in other AFM materials such as $CeAgAs_2$ [72], $EuIn_2As_2$ [73], $Eu_{14}MnBi_{11}$ [74], and $Eu_3Ni_4Ga_4$ [75]. Generally, the negative MR can arise from various mechanisms such as magnetic field-induced modification to electronic band structures [76–78], Kondo effect [79], weak localization [80,81], and chiral anomaly [82–85]. The change in electronic structure should lead to strong modifications to the carrier density, which can be probed by the Hall effect [76,86]. In $\alpha$-$Gd_2Se_3$, however, as shown in the inset of Fig. 3(c), the Hall resistivity $\rho_{yx}(H)$ does not exhibit a strong deviation from a linear field, implying an almost unchanged electronic structure under a magnetic field. Furthermore, the carrier density extracted from the slope of $\rho_{yx}(H)$ displays rather weak temperature dependence from 2 to 300 K [Fig. 3(c)], suggesting that the AFM transition may not notably change the band structure. Near zero field, $\rho_{yx}(H)$ displays very weak non-linearity below 20 K. Since Magnetization and $\rho_{xy}$ under $H\perp(201)$ do not evolve with the magnetic field coincidently, such nonlinearity is less likely to originate from the anomalous Hall effect. Hence it is better attributed to a multi-band effect. In general, the multi-band effect is manifested in both longitudinal $\rho_{xx}$ and transverse ($\rho_{yx}$) resistivity, producing a nearly $H^2$-like field dependence for $\rho_{xx}(H)$ and nonlinearity in $\rho_{yx}(H)$. Providing the existence of multiple correlated fitting parameters, the carrier density and mobility for each band should be obtained via simultaneous fitting of both $\rho_{xx}(H)$ and $\rho_{yx}(H)$ to the multi-band model. However, given longitudinal resistivity $\rho_{xx}(H)$ for $\alpha$-Gd$_2$Se$_3$ exhibits a negative MR that is not described by the multi-band model, it is thus not possible to obtain reliable carrier densities and mobilities. Nevertheless, providing that the nonlinearity in $\rho_{yx}(H)$ is rather weak, electronic transport in $\alpha$-Gd$_2$Se$_3$ is dominated by one band. Therefore, the carrier density $n$ can be estimated from the single-band model as shown in Fig. 3(c), from which the carrier mobility can be calculated via $\mu = 1/(ne\rho_{xx})$, as presented in Fig. 3(d).

Similarly, the Kondo effect due to the screening of dilute magnetic moments by carriers can also be ruled out, because it should lead to low-temperature resistivity upturn with a logarithmic temperature dependence [79,87–89] which is not observed in $\alpha$-Gd$_2$Se$_3$. In addition, the observed large negative MR is much higher than those observed for the Kondo system [89–91]. The weak localization can be excluded as well. This effect is caused by the enhanced backscattering rate due to the constructive interference of the time-reversal backscattering carrier paths. Applying a magnetic field suppresses the quantum interference and lowers the backscattering rate, leading to negative MR [80,81]. However, MR is expected to saturate quickly with the magnetic field at low temperatures owing to the efficient suppression of quantum interference, which is not observed in $\alpha$-Gd$_2$Se$_3$. The negative MR in $\alpha$-Gd$_2$Se$_3$ should not be ascribed to chiral anomaly either. This phenomenon, i.e., imbalance of chiral fermions, arises from the charge pumping between a pair of Weyl cones under parallel electrical and magnetic fields [82,83], and hence it is sensitive to the directions of the magnetic field. Fig. 4(a) shows the MR at 2 K measured at various field orientations. Strong MR ~ 54% at 9 T can be observed for all field orientations from $H // I$ ($\theta = 90°$) and $H \perp I$ ($\theta = 0°$), indicating chiral anomaly is not applicable.

With ruling out the other mechanisms, the most likely origin for the strong negative MR in $\alpha$-Gd$_2$Se$_3$ is the suppression of magnetic scattering by field. Our transport measurements reveal a very strong interplay between magnetism and electronic transport. As stated above, the temperature dependence for resistivity at zero magnetic field displays a sharp peak at $T_N$ ~ 11 K [Fig. 3(a)]. This can be understood in terms of the enhanced magnetic scattering near the magnetic ordering temperature where the spin fluctuations are the strongest. In fact, the temperature dependent resistivity at zero magnetic field starts to develop an upturn at 55 K, which is much higher than $T_N$ and implies sizeable magnetic scattering above $T_N$. Such a scenario is also consistent with the strong field-induced heat capacity enhancement above $T_N$ mentioned above [Fig. 2(f)]. Therefore, resistivity reduction due to the suppression of magnetic fluctuations by the magnetic field is expected. Such suppression should be the most significant at $T_N$ and consequently leads to the strongest negative MR near $T_N$ as observed in our magnetotransport measurements [Figs. 3(b) and 3(c)]. Above $T_N$, MR is also reduced upon heating because the thermal energy randomizes magnetic moment orientations, and it becomes more difficult to polarize these moments by the magnetic field. However, providing strong magnetic correlations up to 55 K as discussed above, MR retains a remarkable value of 46% at 20 K and remains 12% at 50 K. The persistence of substantial magnetic correlations well above the magnetic transition temperature in Gd$_2$Se$_3$ appears to be consistent with other magnetic compounds containing Gd such as GdPS [59]. Additionally, temperature dependent mobility also provides support to the scenario of magnetic scattering. As shown in Fig. 3(d), at low temperatures, mobility $\mu$ increases because of suppressing magnetic fluctuations. $\mu$ reaches a local minimum around $T_N$ due to the strong spin scattering with the presence of intense magnetic fluctuations. With increasing the temperature, the spin scattering is suppressed but the electron-phonon interaction becomes strengthened. The competition of the two mechanisms leads to non-monotonic temperature dependence for mobility for $T > T_N$. Mobility slightly enhances up to ~50 K where spin fluctuations start to develop, while drops at higher temperature when electron-phonon interaction dominates.

In addition to the temperature dependence, the field dependence of resistivity also provides additional support. As shown in Fig. 3(b), for $H\perp(201)$, though MR at 2 K is lower than that at $T_N$, its saturation behavior resembles that of the magnetization saturation [Fig. 2(c)]. This should be attributed to the nearly complete suppression of spin fluctuations when magnetic moments are fully polarized. Similar low-temperature saturation behavior in MR [Fig. 4(a)] and magnetization [Fig. 2(d)] is also observed under the in-plane $H//(201)$ field. The multiple peak-like features in MR for $H//(201)$ in Fig. 4(a) should be ascribed to metamagnetic SF transitions. Figure 4(b) presents the field dependence for magnetization and MR under $H\perp(201)$ and $H//(201)$ at 2 K. The metamagnetic transitions at 0.85 T and 2.35 T for $H//(201)$ are accompanied by positive MR whereas the metamagnetic transition at 3.55 T is too weak to develop a clear positive MR but rather exhibits a weak slope change in MR. On the other hand, for $H\perp(201)$ where the metamagnetic transition is not present, MR lacks any other feature exhibit high field saturation. These observations can be understood in terms of the spin scattering. Upon applying a magnetic field near an SF field, the spin scattering is strong due to strong spin fluctuations in the vicinity of the SF transition. Further, increasing the field suppresses spin fluctuations and reduces scattering. Hence a peak-like feature and positive MR is observed near SF transition fields.

The angular MR (AMR) is also consistent with the scenario of spin scattering. Figures 5(a)and 5(b) show the angular dependence for resistivity measured at fixed magnetic fields from 1 to 9 T at $T = 2$ K (AFM state) and 15 K (PM state), respectively. At 2 K, the low field (1 and 2 T) AMR displays relatively complicated angular dependence with multiple peaks, with an overall two-fold anisotropy with the maxima and minima at $H//(201)$ and $H\perp(201)$, respectively. Such complicated MR anisotropy should be caused by the multiple metamagnetic transitions that are sensitive to magnetic field orientation as seen in the field-dependent MR in Fig. 4(a). The low field MR reaches a maximum when the field is applied along the magnetic easy axis [i.e., $H//(201)$], which can be understood in terms of enhanced spin scattering at SF transition as discussed above. With an increasing magnetic field, an AMR dip at $H//(201)$ starts to develop, causing a four-fold-like AMR anisotropy at 3T. Such AMR dip becomes more significant with further increasing magnetic field, leading to AMR minima at $H//(201)$ above 5 T, as shown in Fig. 5(a). This can be attributed to the strong suppression of spin scattering when $H//(201)$. Unlike the perpendicular field $H\perp(201)$, the in-plane field $H//(201)$ induces SF transitions with which the FM component develops more rapidly, leading to a strong increase and quick saturation of magnetization at higher fields. Therefore, the spin scattering is more significantly suppressed for $H//(201)$ at higher fields, causing the AMR to change anisotropy with the field. At temperatures above $T_N$, a similar two-fold AMR anisotropy with minima at $H//(201)$ remains observable at 15 K for various applied magnetic fields (Fig. 5b), which agrees with the presence of strong magnetic correlations above $T_N$ mentioned above and further supports the scenario of negative MR due to spin scattering.

## IV. CONCLUSION

In conclusion, we have successfully synthesized the single crystal of the orthorhombic phase of $Gd_2Se_3$ by chemical vapor transport and studied its transport, magnetic, and calorimetric properties. We found that $\alpha$-$Gd_2Se_3$ possesses an AFM order below $T_N \approx 11$ K which can be driven into a polarized FM state at higher fields. In the PM state, magnetic fluctuations remain strong. The transport measurements reveal metallic-like behavior and large negative MR near $T_N$, which should be attributed to the suppression of spin scatterings. Overall, $\alpha$-$Gd_2Se_3$ behaves as an intermediate material between non-metallic $\alpha$-$Gd_2S_3$ and metallic α-Gd$_2$Te$_3$ in terms of magnetic properties, displaying strong modification of electron transport by magnetism. Therefore, magnetism and transport is expected to be highly tunable by various approaches such as chemical substitution, pressure, and strain, which might provide a versatile platform for spintronics applications.

**Acknowledgement**

This work was primarily (synthesis, heat capacity, magnetization up to 9T, and transport) supported by the U.S. Department of Energy, Office of Science, Basic Energy Sciences program under Grant No. DE-SC0022006. M. M. and R. B. acknowledge µ-ATOMS, an Energy Frontier Research Center funded by DOE, Office of Science, Basic Energy Sciences, under Award DE-SC0023412 for part of magnetism and structural analyses. J. S. acknowledges the support from NIH under award P20GM103429 for XRD. J. H. acknowledges the MonArk NSF Quantum Foundry supported by the National Science Foundation Q-AMASE-i program under NSF award No. DMR-1906383 for 7T magnetization study using MPMS3 SQUID. High field magnetotransport was performed at the National High Magnetic Field Laboratory, which is supported by National Science Foundation Cooperative Agreement No. DMR-2128556 and the State of Florida.

**Figures**

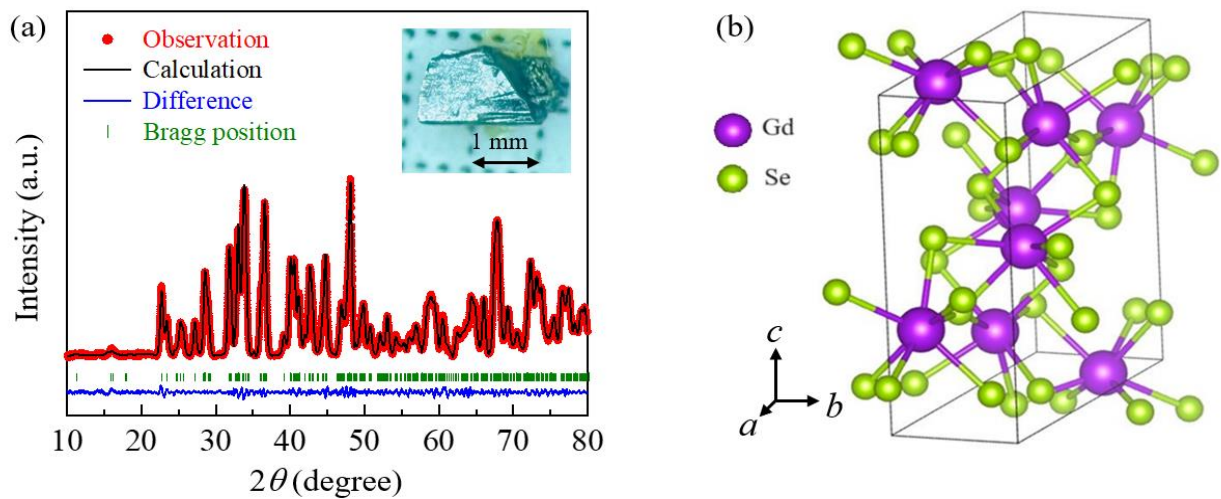

FIG. 1. (a) Powder XRD data and the Rietveld refinement of α-Gd$_2$Se$_3$. Inset: image of a α-Gd$_2$Se$_3$ single crystal. (b) Crystal structure of the orthorhombic α-Gd$_2$Se$_3$ obtained from the refinement. The structure parameters are provided in Table I.

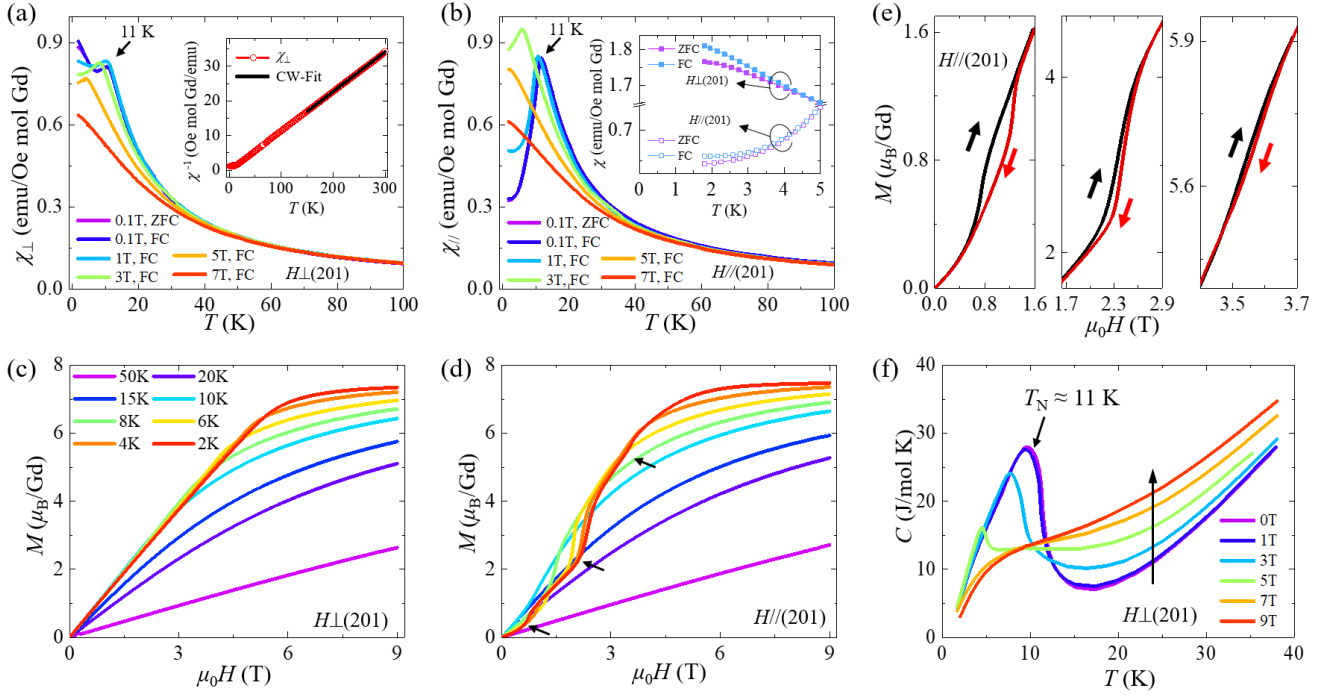

FIG. 2. Magnetic properties of α-Gd$_2$Se$_3$. (a) Temperature dependence of molar susceptibility of α-Gd$_2$Se$_3$ measured under $H\perp(201)$ magnetic fields from 0.1 to 7 T. Inset: CW fit for the inverse susceptibility measured at 0.1 T. (b) Temperature dependence of molar susceptibility measured under various $H//(201)$ magnetic fields from 0.1 to 7 T. Inset: zoom-in of ZFC and FC susceptibility below 5 K measured with 0.1 T field. (c-d) Field dependence of magnetization with (c) out-of-plane $H\perp(201)$ and (d) in-plane $H//(201)$ magnetic fields at different temperatures. The same color code is used for (c) and (d) to distinguish each temperature. Arrows in (d) indicate metamagnetic transitions. Magnetic hysteresis of these metamagnetic transitions at 2 K are shown in (e). (f) Temperature dependence of heat capacity of α-Gd$_2$Se$_3$ measured under various magnetic fields applied along the out-of-plane [$H\perp(201)$ plane] direction from 0 to 9 T.

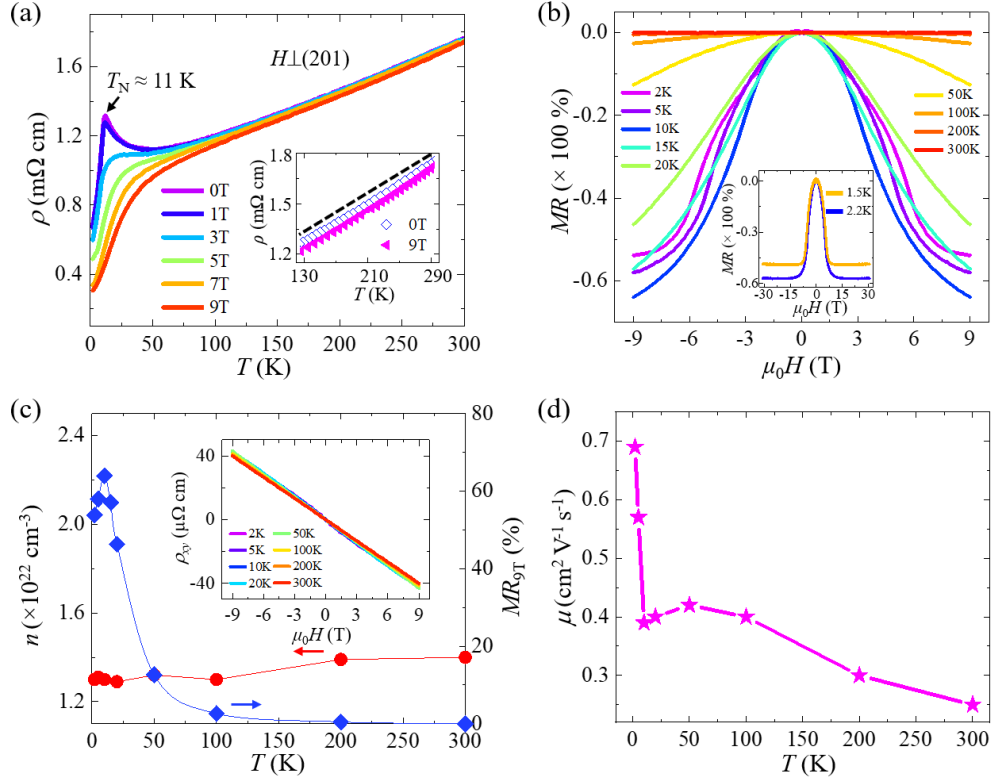

FIG. 3. Magnetotransport properties of $Gd_2Se_3$. (a) Temperature dependence of resistivity of $Gd_2Se_3$ under various magnetic fields applied perpendicularly [$H\perp(201)$]. Inset: linear temperature dependence for resistivity at high temperatures under 0 T and 9 T fields. The black dashed lines are guides for the eyes. (b) Normalized MR at different temperatures. Inset: High field MR measured up to 31 T at 1.5 K and 2.2 K. (c) Temperature dependence of carrier concentration extracted from Hall effect ($n$, left vertical axis) and magnitude of MR at 9 T field ($MR_{9T}$, right vertical axis). Inset: Field dependence of Hall resistivity at different temperatures. (d) Temperature dependence of carrier mobility $\mu$ extracted using the single-band model.

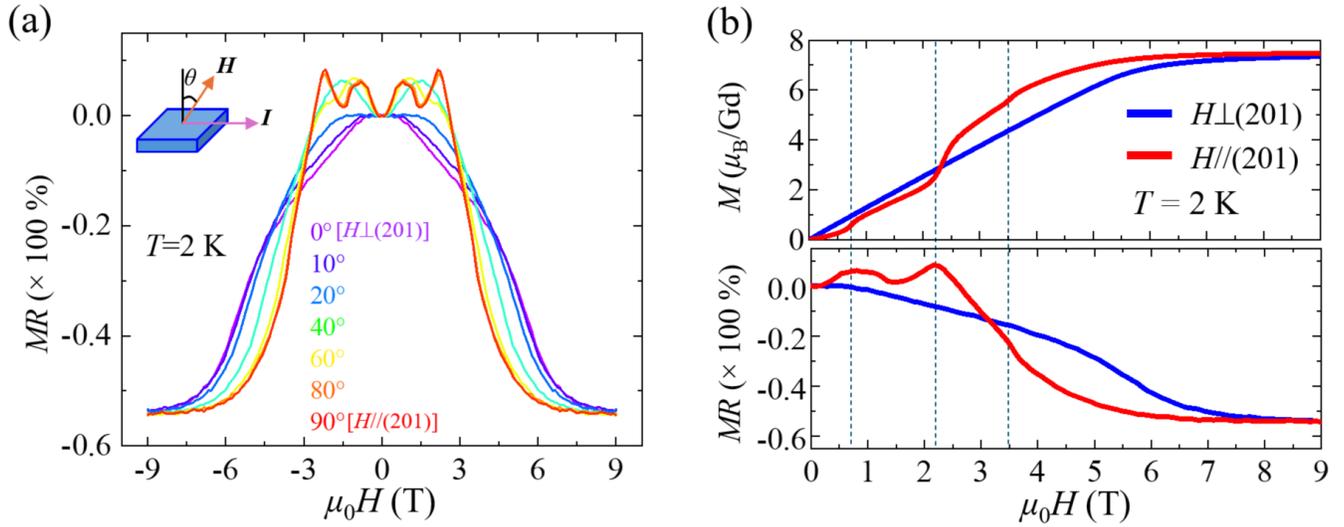

FIG. 4. (a) MR of Gd$_2$Se$_3$ at $T$ = 2 K under different magnetic field orientations. The measurement setup is shown in the Inset. (b) Comparison of the field-driven transitions between magnetization (upper panel) and MR (lower panel) under $H\perp(201)$ and $H//(201)$ fields at $T$ = 2 K, which are reproduced from Figs. 2(c), 2(d), and 4(a). The vertical dashed lines denote the metamagnetic transition fields.

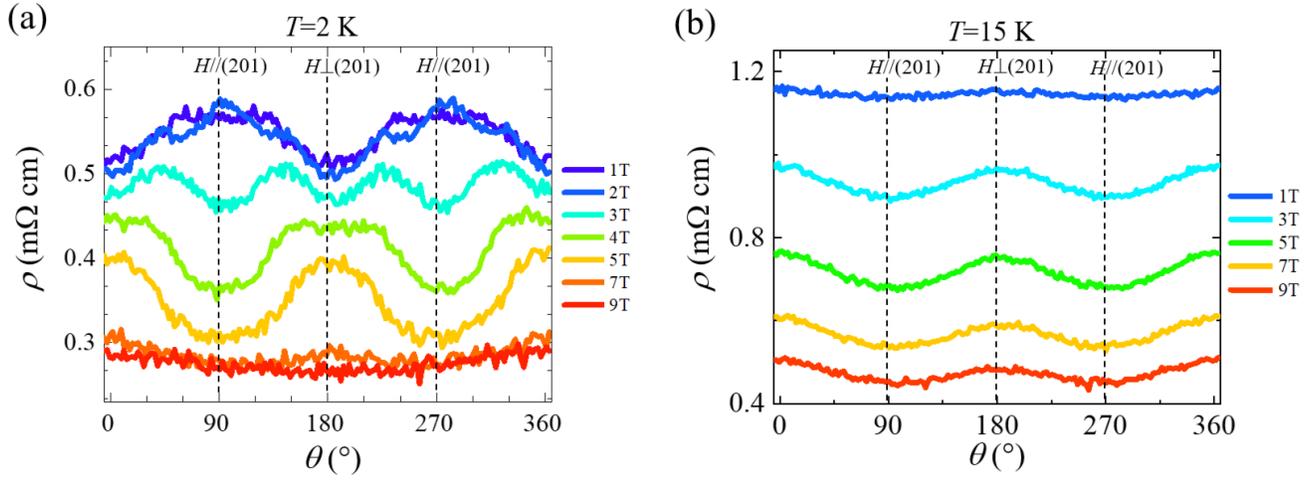

FIG. 5. Angular dependence of resistivity of α-Gd$_2$Se$_3$ single crystal below (2 K, panel a) and above (15 K, panel b) $T_N$ measured under different magnetic fields. The dashed lines denote $H//(201)$ and $H\perp(201)$ field orientations

**Table**

Table I. Structural parameters of $Gd_2Se_3$ at T = 300 K. Space group: *Pnma*; $a$ = 11.177(1) Å, $b$ = 4.049(4) Å, $c$ = 10.966(4) Å; $\alpha = \beta = \gamma = 90°$, $R_p$ = 4.87, $R_{wp}$ = 4.64.

| Atoms | Wycoff | $x$ | $y$ | $z$ |
|---|---|---|---|---|
| **Gd1** | 4c | 0.9901(3) | 1/4 | 0.3117(5) |
| **Gd2** | 4c | 0.3046(4) | 1/4 | 0.5042(1) |
| **Se1** | 4c | 0.0453(4) | 1/4 | 0.8753(3) |
| **Se2** | 4c | 0.8763(8) | 1/4 | 0.5585(9) |
| **Se3** | 4c | 0.2259(9) | 1/4 | 0.1956(6) |